\documentclass[amsmath,amssymb,reprint,aps,prx,floatfix,superscriptaddress]{revtex4-2}
\usepackage{bm}
\usepackage{amsmath}
\usepackage{graphicx,float}
\usepackage{braket}
\usepackage{bbm}
\usepackage[hidelinks]{hyperref}
\usepackage[inline]{enumitem}
\usepackage[qm]{qcircuit}
\usepackage{xcolor} %SK added for coloured text
\begin{document}

\preprint{APS/123-QED}

\title{Experimentally finding dense subgraphs using a time-bin encoded Gaussian boson sampling device}

\author{S. Sempere-Llagostera}
%\thanks{These two authors contributed equally.}
\affiliation{Department of Physics, Imperial College London, Prince Consort Rd, London SW7 2AZ, UK}

\author{R. B. Patel}
\affiliation{Department of Physics, Imperial College London, Prince Consort Rd, London SW7 2AZ, UK}
\affiliation{Clarendon Laboratory, University of Oxford, Parks Road, Oxford, OX1 3PU, UK}

\author{I. A. Walmsley}
\affiliation{Department of Physics, Imperial College London, Prince Consort Rd, London SW7 2AZ, UK}

\author{W. S. Kolthammer}
\affiliation{Department of Physics, Imperial College London, Prince Consort Rd, London SW7 2AZ, UK}

\date{\today}% It is always \today, today,

\begin{abstract}
Gaussian boson sampling (GBS) is a quantum computing concept based on drawing samples from a multimode nonclassical Gaussian state using photon-number resolving detectors. %Unlike other near-term protocols aiming to achieve quantum advantage, 
It was initially posed as a near-term approach to achieve quantum advantage, and several applications have been proposed since, including the calculation of graph features. For the first time, we use a time-bin encoded interferometer to implement GBS experimentally and extract samples to enhance the search for dense subgraphs in a graph. Our results indicate an improvement over classical methods for subgraphs of sizes three and four in a graph containing ten nodes. In addition, we numerically explore the role of imperfections in the optical circuit and on the performance of the algorithm.
\end{abstract}

\maketitle

%\tableofcontents

\section{\label{sec:Introduction}Introduction}
Quantum computing promises to modify the current computational paradigm~\cite{Michael2010, Ladd2010} by providing a substantial speedup in the computation of specific tasks that are currently intractable. Two key milestones achieved so far are \begin{enumerate*}[label = (\roman*)] \item theoretically developing algorithms that show a quantum speedup compared to the classical counterpart~\cite{Shor2006}, and \item experimentally implementing protocols that surpass the capabilities of traditional computers~\cite{Arute2019, Zhong2021QuantumPhotons} \end{enumerate*}. Meeting these two at a common ground to show a quantum advantage in practical applications is a compelling next step. Progress towards a fault-tolerant universal quantum computer brings us closer to this goal, yet it is possible that a shortcut is provided by simpler, limited-purpose computers, perhaps without error-correction~\cite{Saggio2021}. %Two main avenues exist to achieve this, either by building a fault-tolerant universal quantum computer or by coming up with devices that can outperform the classical method in a specific task conceived as useful. Despite consistent progress over the last few years, the former still seems a bit far away in time, while the latter exhibits promising prospects in the near term for particular scenarios~\cite{Saggio2021}.

In the optical domain, Gaussian boson sampling~\cite{Hamilton2017, Kruse2019} has emerged as a potential candidate in this regard. GBS involves injecting a multimode nonclassical Gaussian state into an interferometer and measuring the number of photons at each output. The classical procedure to simulate this sampling problem is not efficiently computable on a classical computer~\cite{Hamilton2017, Aaronson2011a}. Specifically, it requires the computation of a mathematical function called the hafnian, which constitutes a \#P-hard problem. So far, GBS has found applications both as a subroutine in the bigger quantum computer picture as a resource state generator~\cite{Tzitrin2020a, EliBourassa2021}, and as a stand-alone device that can be used, for instance, to calculate vibronic spectra of
molecules or graph features~\cite{Huh2015BosonSpectra, arrazola2018dense, Bromley2019}. Concerning the two key aforementioned milestones, a quantum advantage has been achieved by means of a GBS experiment~\cite{Zhong2021QuantumPhotons, zhong2021phase}, but no practical application has been demonstrated yet. However, in a recent paper, Arrazola et al.~\cite{Arrazola2021QuantumChip} showed promising results employing a fully reconfigurable small scale 8-mode GBS device for the calculation of molecular vibronic spectra and graph similarity. 

A technical challenge of optical interferometers at the core of GBS is to scale them up to a large number of modes, while achieving low loss and reconfigurability. Traditionally, the information carried by an optical state is encoded in its polarisation or spatial modes, implying the need for extensive hardware requirements when employing a few dozens of photons. As shown by Reck et al.~\cite{Reck1994}, the number of beam splitters needed to implement an arbitrary $m\times m$ unitary matrix is $m(m-1)/2$. In the context of quantum information processing and GBS, these have been implemented in various forms, for instance, using bulk optics \cite{He2017} and in integrated platforms \cite{Spring2013, Anton2019, Taballione2020AProcessing}. Other non-universal approaches have been achieved using bulk optics~\cite{Wang2019} and fibre architectures~\cite{Boutari2016, Broome2013}, the largest to date being a bulk optics interferometer with 144 input/output modes~\cite{zhong2021phase}. %Ideally, an interferometer should be lossless, reconfigurable and scalable in the number of input and output modes. 
A recent proposal~\cite{Motes2014} suggested using the temporal modes to encode the information as a means to scale up for boson sampling experiments. This scheme is attractive for its scalability and reconfigurability \cite{He2017}, whilst exhibiting comparable losses \cite{Qi2018a, Motes2015a} to the other reconfigurable platforms \cite{Taballione2020AProcessing}. Furthermore, in this approach, the number of components is fixed regardless of the size of the implemented unitary. In principle, arbitrary scale can be achieved with minor hardware modifications. Alternative non-universal time-bin interferometers have been proposed to achieve quantum supremacy via high-dimensional GBS~\cite{Deshpande2021}.

%In this work, we harness the scalability of the time-bin interferometer to experimentally show how a GBS device can enhance the search for dense subgraphs. In particular, we demonstrate this for subgraphs of sizes three and four in a graph of ten nodes. This constitutes the first experimental realisation of GBS in a time-bin encoded manner and the first practical usage of a GBS machine to improve the search of dense subgraphs. 
In this work, we \begin{enumerate*}[label = (\roman*)] \item implement a 20-mode time-bin interferometer to conduct a GBS experiment, and \item experimentally show how such GBS device can enhance the search for dense subgraphs \end{enumerate*}. In particular, we demonstrate this for subgraphs of sizes three and four in a graph of ten nodes. This constitutes the first experimental realisation of GBS in a time-bin encoded manner and the first application of a GBS machine to improve the search of dense subgraphs. 

The paper is structured as follows. First, we introduce the main concepts of Gaussian boson sampling, the time-bin interferometer, and the use of GBS to search for dense subgraphs in a graph. Then, we describe the experimental work, which combines these three elements, and show the results obtained. Finally, we perform numerical calculations to study the role of imperfections for the particular application of finding dense subgraphs. %Finally, we summarise our results and give an outlook of the next steps needed to use GBS to provide an absolute speed up for the problem of finding dense subgraphs.
\section{\label{sec:Theory}Theory}
\subsection{\label{sec:GBSTime} Gaussian boson sampling}
Figure \ref{fig:concept}(a) illustrates the concept of Gaussian boson sampling. We consider an $m$-mode system with separable Gaussian state inputs that undergoes a transformation characterised by the transfer matrix $\Lambda$, followed by single-photon detection. The input states, $|\xi_i\rangle$, can be chosen from the spectrum of Gaussian states, for instance, vacuum, squeezed vacuum, or coherent states, among others. The inteferometer only consists of linear optical elements and loss, which are Gaussian transformations~\cite{Weedbrook2011}. Therefore, the state at the output is also Gaussian and can be completely described by a $2m\times2m$ covariance matrix, $\bm{\Sigma}$, and a displacement vector, $\bm{\alpha}$. With no displacement, i.e. $\bm{\alpha} = 0$, the probability to measure a certain output pattern $\bm{n} = (n_1, n_2, n_3, \dots, n_m)$ is given by \cite{Hamilton2017}
\begin{equation}
\label{eq:p_n}
    P(\bm{n})=\frac{P(0)}{\prod_i n_i!}\mathrm{Haf}(\bm{A_\bm{n}}),
\end{equation}
where $P(0)$ is the probability of not detecting any photons at the output and $\mathrm{Haf}$ is the hafnian function. The hafnian of a matrix is computed as 
\begin{equation}
\label{eq:haf}
    \mathrm{Haf}(\bm{A}) =\sum_{M \in \operatorname{PMP}(n)} \prod_{(i, j) \in M} A_{i, j} \,
\end{equation}
where $\mathrm{PMP}$ indicates the set of perfect matching permutations of $n$ objects. The matrix $\bm{A}$ can be derived from the covariance matrix as 
\begin{equation}
    \bm{A} = \bm{X}(\bm{I}-\bm{Q}^{-1}),
\end{equation}
where 
\begin{subequations}
\begin{align}
   \bm{X} = \begin{pmatrix}
0 & \bm{I}\\
\bm{I} & 0 
\end{pmatrix}\quad \mathrm{and}  \label{eqn:X} \\
    Q = \bm{\Sigma}+\bm{I}/2 \label{eqn:Q},
\end{align}
\end{subequations}
where $\bm{I}$ is the identity matrix of appropriate dimension. The submatrix $\bm{A_\bm{n}}$ is obtained by selecting $n_i$ times the $i$-th and $i+m$-th rows and columns of $\bm{A}$. Consequently, $\bm{A_\bm{n}}$ is a square matrix that grows with the total number of detected photons $N$, with $N = \sum_i n_i$. The matrix $\bm{A}$ can be decomposed into a block matrix of the form, 
\begin{equation}
\label{eq:Adecompos}
    \bm{A} = \begin{pmatrix}
\bm{B} & \bm{C}\\
\bm{C^\mathrm{T}} & \bm{B^*} 
\end{pmatrix},
\end{equation}
where $\bm{B}$ is a $m\times m$ symmetric matrix and  $\bm{C}$ is a $m\times m$ Hermitian matrix. 

In the particular case of having a pure Gaussian state with no displacement at the output, $\bm{A}$ can be further reduced to $\bm{A} = \bm{B}\oplus\bm{B^*}$, i.e. $\bm{C} = 0$, and the probability to measure a detection pattern $\bm{n}$ becomes
\begin{equation}
\label{eq:haf_B}
    P(\bm{n})= \frac{P(0)}{\prod_i n_i!}|\mathrm{Haf}(\bm{B_\bm{n}})|^2,
\end{equation}
where the submatrix $\bm{B_\bm{n}}$ is obtained by selecting $n_i$ times the $i$-th rows and columns. Note that in this case, the matrix $\bm{B_\bm{n}}$ will be of size $N\times N$ and, due to the nature of the hafnian, $N$ needs to be even. This agrees well with the fact that if there is no loss present in the system, which would add mixedness to the state, photons will always come in pairs from the squeezed vacuum states and, therefore, detecting an odd number of photons will not be possible. 

\subsection{\label{subsec:timebinint}Time-bin interferometer}
\begin{figure}
    \centering
    \includegraphics[width=1\columnwidth]{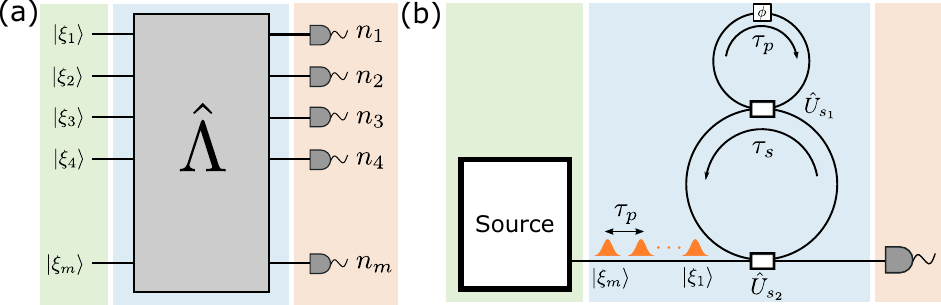}
    \caption{(a) General GBS picture, where we have an $m$-mode Gaussian state going into an interferometer described by $\hat{\Lambda}$ and is then detected using single-photon detectors. (b) Time-bin encoded architecture to implement GBS experiments. The modes are defined by temporal bins, and the interferometer is implemented by the double-loop structure. One light source generates the multimode input state, which is detected at the output by one single-photon detector.}
	\label{fig:concept}
\end{figure}

Here we introduce the time-bin interferometer~\cite{Motes2014}, motivate its usage and explain how it can implement an arbitrary unitary. Fig.~\ref{fig:concept}(b) exhibits an abstract representation of the end-to-end proposed architecture. As usual, we can distinguish three main parts, the light source, the interferometer and the detection scheme, interfaced via a common spatial mode. 

%We have labelled the source as a source of single-mode squeezed vacuum (SMSV), but it can be generalised to any type of light source. 
%The light source generates a train of pulses with separation $\tau_p$, which defines the time-bin modes and contains the input states $\ket{\xi_i}$. The fact that we only need a single source to run the whole experiment, as opposed to many sources \cite{Spring2013, Zhong2021QuantumPhotons} or de-multiplexing techniques \cite{Anton2019, Wang2019}, poses a significant simplification in terms of experimental resources. The laser system pumping the source will define the different temporal modes, meaning that to enlarge the size of the problem, we only need to let more pulses into the interferometer. For instance, a Ti:Sapphire laser with a repetition rate on the order of a few MHz and pulses of picoseconds could be employed.

The light source generates a train of pulses with separation $\tau_p$, which defines the time-bin modes and contains the input states $\ket{\xi_i}$. The fact that we only need a single source to run the whole experiment, as opposed to many sources \cite{Spring2013, Zhong2021QuantumPhotons} or de-multiplexing techniques \cite{Anton2019, Wang2019}, poses a significant simplification in terms of experimental resources. In addition, enlarging the size of the problem is straightforward, as it solely requires letting more pulses into the interferometer. %For instance, a Ti:Sapphire laser with a repetition rate on the order of a few MHz and pulses of picoseconds could be employed.

The interferometer consists of two loops arranged in a snowman configuration, the lower one is the ``storage'' loop and the upper one is the ``processing'' loop. The length of the processing loop is matched to the separation between consecutive pulses $\tau_p$, while the storage loop must accommodate all the interfering $m$ modes, i.e. $\tau_s \ge m\tau_p$. The two loops are connected via a variable beam splitter, s$_1$, and to the input/output mode through a fast switch, s$_2$. %Note that s$_2$ has a dual purpose: coupling the input photons into the interferometer and setting the remaining modes to vacuum, i.e. acting as a shutter. 
The main distinction between s$_1$ and s$_2$ is the number of possible reflectivities each of them needs to be able to achieve. While s$_2$ only acts as a switch, so it has an on/off behaviour, s$_1$ is the beam splitter that will implement the unitaries and, therefore, needs to be able to cover a large range of possible splitting ratios. 

\begin{figure}
    \centering
    \includegraphics[width=1\columnwidth]{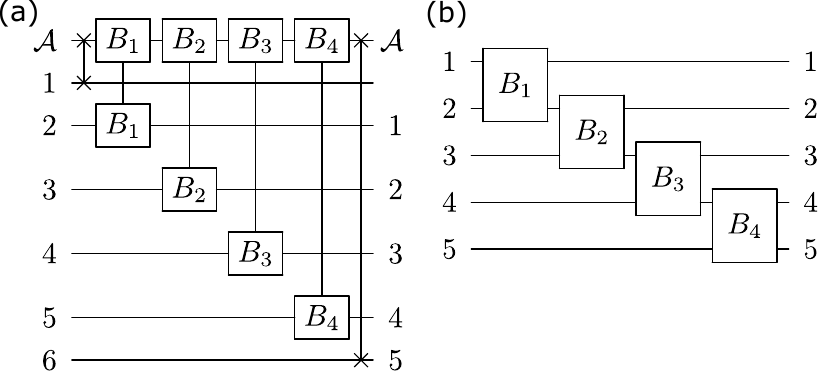}
    \caption{(a) Explicit circuit of a pass through the processing loop. As per convention, the x symbol refers to the SWAP operation. (b) Simplified circuit of a single pass through the processing loop. The mode labels refer to time-bin modes, except for $\mathcal{A}$ that is the processing loop mode, and $B_i$ represents beam splitter operation.}
	\label{fig:circuit}
\end{figure}

Here we show how this approach can implement any multimode interferometer \cite{Motes2014, Qi2018a}. The basic concept is that the processing loop enables the interference of consecutive modes, whilst the storage loop enables this to be repeated as many times as required. The circuit in Fig.~\ref{fig:circuit}(a) describes one round trip of the storage loop, during which each pulsed mode in the storage loop, denoted by a numbered line, interacts sequentially with a pulsed mode in the processing loop, denoted by the line $\mathcal{A}$.
The coupling between the loops is described by $B_i$, corresponding to a general two-mode beam-splitting operation -- a phase shifter followed by a Mach-Zehnder interferometer, for instance.
In this example, the initial state is contained in modes 1-5, while modes $\mathcal{A}$ and 6, the last storage mode, are initialized as vacuum states. The SWAP operations at the start and end of the sequence leave modes $\mathcal{A}$ and the first storage mode, as vacuum states. At the output of Fig.~\ref{fig:circuit}(a), the modes are relabeled so that those containing the processed state are consistently numbered 1-5.
%The SWAP operation couples the corresponding mode into the loop mode, $\mathcal{A}$.

In Fig.~\ref{fig:circuit}(b), the auxiliary vacuum modes are disregarded, and one round trip of the storage loop is seen to enact sequential beam-splitting operations between consecutive modes. This process is repeated by choosing appropriate on/off timing for s$_2$. Notably, a series of $m-1$ round trips can be used to implement the $m$-mode interferometer decomposition by Reck et al. \cite{Reck1994}, and therefore the snowman structure is universal for linear optics.
Similar approaches may be used to implement arbitrary interferometers using other decompositions~\cite{ Clements2018a, Clements2016OptimalInterferometers}, which might offer practical improvements in mitigating non-uniform loss incurred across different input-output pairs~\cite{Motes2015a}. Lastly, we note that universality may still be achieved for limited beam-splitting values \cite{Bouland2014a, Sawicki2016UniversalityBeamsplitters}, at the cost of additional round trips and corresponding optical loss.

Finally, we have the detection system. This presents another substantial improvement of this architecture over other platforms. Due to the temporal encoding of the information, we can potentially reduce the required number of detectors to one, compared to the spatial encoding depicted in Fig.~\ref{fig:concept}(a) in which we need $m$. This feature can become very relevant when scaling up to bigger interferometers. The main practical consideration is the reset time of the detectors being larger than the time-bin spacing, which could skew the output statistics.

\subsection{\label{sec:densesubgraphs} Finding dense subgraphs using a GBS device}
Here we review a connection between GBS outcome statistics and the identification of highly connected subgraphs, as introduced by Ref.~\cite{arrazola2018dense}.
To start, we note that any graph may be described by its adjacency matrix $\bm{\Delta}$, where the matrix elements $\Delta_{ij}$ determine the weight of the edge between nodes $i$ and $j$. For the case of an undirected graph, $\bm{\Delta}$ is a symmetric matrix, i.e. $\Delta_{ij}=\Delta_{ji}$.

In Eq. \eqref{eq:Adecompos}, we saw that a pure Gaussian state with zero displacement may be described by the symmetric matrix $\bm{B}$. This establishes a correspondence between this quantum state and an undirected graph, in terms of its adjacency matrix. Moreover, the probability of any particular GBS outcome from this state is related to the hafnian of a particular subgraph of $\bm{B}$, which provides information about its connectivity. Notably, the number of terms contributing to the hafnian of a matrix is related to the number of perfect matchings in a graph, as seen in Eq.~\eqref{eq:haf}.

We can define the density of a weighted graph as 
\begin{equation}
    \label{eq:density}
    d = 2\frac{\sum \Delta_{ij}}{|V|(|V|-1)},
\end{equation}
%where $w_i$ are the weights of the edges of $G$, $E_G$, and $|V_G|$ are the number of vertices in $G$. 
where the sum runs over all the edges of the graph and $|V|$ is the number of nodes in the graph, i.e. dim($\bm{\Delta}$). For an unweighted graph, i.e. $\Delta_{ij} \in {0,1}$, the density is bounded between 0 for an unconnected graph and 1 for a fully connected, or complete, graph with $\Delta_{ij}= 1 \ \forall i,j$. It was recently shown by Arrazola and Bromley~\cite{arrazola2018dense}, that the number of perfect matchings is related to the density of a graph, and, therefore, so is the hafnian of a fully positive (or negative) matrix. The higher the number of perfect matchings the denser the graph will be. This means that we can employ a GBS device to generate samples from a certain problem matrix, and that these will form dense subgraphs of the graph described by $\bm{B}$ with high probability. They show that this strategy can enhance the search over uniform random searches or when using these samples as seeds in more elaborate algorithms, for instance, simulated annealing. More generally, this constitutes an example of the enhancement provided by using proportional sampling over random search in specific optimisation problems, as shown in~\cite{arrazola2018quantum}.

The recipe for employing a GBS device for this purpose is given in~\cite{Bromley2019} as follows:
\begin{enumerate}
    \item Decompose $\bm{B}$, which describes the graph we want to study, into a unitary matrix $\bm{U}$ and a vector $\bm{\lambda}$ using the Takagi-Autonne decomposition. These will correspond to the linear interferometer and squeezing parameters at each of the input modes, respectively.
    \item Compile $\bm{U}$ into the appropriate beam splitter and phase-shifting operations of the linear interferometer. Program the interferometer accordingly.
    \item Rescale the squeezing parameters $r_i = \tanh^{-1}(c\lambda_i)$ according to the constant $c>0$ such that $\langle n \rangle=\sum_{i=1}^m\frac{(c\lambda_i)^2}{1-(c\lambda_i)^2}$. In the case of an $n$-node subgraph, this can be used to maximize the probability of obtaining an $n$-fold coincidence at the output. Note that this rescaling does not change the $n$-fold relative probability distribution. Program the squeezers accordingly.
\end{enumerate}
After following this procedure the GBS device will generate sample outcome, $\bm{n}$, with probability 
\begin{equation}
    P(\bm{n}) \propto \frac{c^N}{\prod_i n_i!} |\mathrm{Haf}(\bm{B_n})|^2
\end{equation}
where $N=\sum_i n_i$ is the total number of detected photons.% and we assume that we are in the collision-free regime, i.e. $n_i = 0,1 \ \forall i$. 
We can then feed these samples into a classical algorithm to find dense subgraphs in a graph, for instance, as shown in \cite{arrazola2018dense}.

An unavoidable imperfection in optical systems is loss. In experiments involving single-photon states, as is the case of standard boson sampling, uniform loss in the interferometer can be neglected by post-selecting on the same number of photons as we had at the input. So, in this scenario, loss will ultimately affect our detection rates solely, but the probability distribution at the output for some $n$-fold detection will remain unaffected. The situation is rather different when employing other types of input states, for instance, single-mode squeezed vacuum (SMSV). In the case of GBS, it is easy to see that if we have loss in the system, we will not have a pure state anymore, and, therefore, $\bm{C}\ne0$ and we would be sampling from a different distribution. The impact on how this imperfection affects our ability to use GBS to find dense subgraphs in a graph is studied later in Sec.~\ref{sec:imperfections}.

\section{\label{sec:Experiment}Experiment}
\begin{figure}[t]
    \centering
    \includegraphics[width=1\columnwidth]{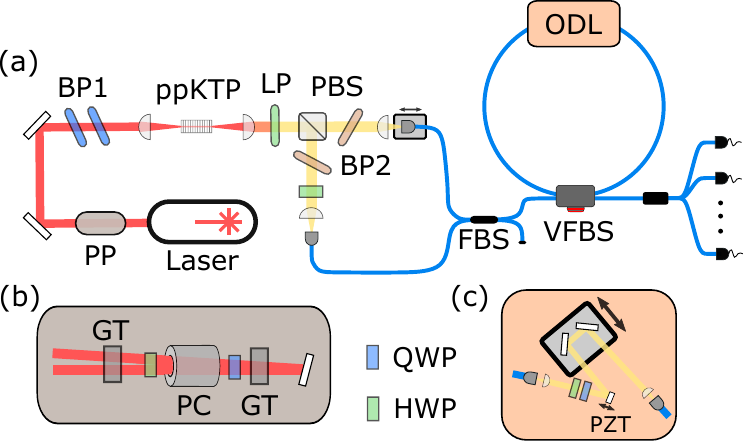}
	\caption{(a) Main experiment setup. PP: pulse-picker, BP1: bandpass filters at 775 nm, PBS: polarising beam-splitter (Wollaston prism), BP2: bandpass filters at 1550 nm, FBS: fibre beam splitter, VFBS: variable FBS and ODL: optical delay line. (b) Detailed schematic of the pulse-picking system, indicated as PP in (a). GT: Glan-Taylor polariser, PC: Pockels cell, QWP: quarter-wave plate and HWP: half-wave plate. (c) Detailed schematic of the free-space optical delay in the single-loop interferometer, indicated as ODL in (a). PZT: piezoelectric actuator.}
	\label{fig:experiment_setup} 
\end{figure}

%Fig.~\ref{fig:experiment_setup} shows a schematic of the setup employed in the experiment. 
Our implementation of time-bin encoded GBS is depicted in Fig.~\ref{fig:experiment_setup}. A mode-locked Ti:Sapphire laser produces 100-fs pulses at a repetition rate of 80 MHz. %A Pockels cell-based pulse picker is used to define the temporal modes of interest and reject groups of pulses in order to initialise the loop to vacuum. 
A Pockels cell-based pulse picker acts as a shutter to create isolated pulse sequences that correspond to individual trials. The off time between trials ensures the initialisation of the loop to vacuum. A pair of angle-tuned bandpass filters shape the spectrum of the incoming pump pulses appropriately to obtain degenerate and factorable emission via spontaneous parametric down-conversion (SPDC) from a type-II periodically poled potassium titanyl phosphate (PPKTP) waveguide~\cite{Bell2019, thekkadath2021measuring, Sempere-Llagostera2022}. The two orthogonally polarised fields of the approximate two-mode squeezed vacuum (TMSV) state generated in the nonlinear process are split using a Wollaston prism and coupled into polarisation-maintaining (PM) single-mode fibre. Rotating the polarisation of one of the fields and using a fibre-based PM 50:50 beam splitter, we recombine the two fields to generate approximate SMSV states. In this case, the two output fields do not exhibit any correlation and we can block one of them and use the other for the time-bin interferometer, preserving its purity. We send one of the outputs to the single-loop interferometer, which consists of an evanescent-field variable fibre coupler and a fibre-based loop. The length of the loop is matched to the repetition rate of the laser by using a free-space optical delay with a motorised stage, as shown in Fig.~\ref{fig:experiment_setup}(c). %The phase of the loop is controlled using a mirror mounted on a piezoelectric actuator.
Fine control of the loop length, which determines the phase, is achieved by using a mirror mounted on a piezoelectric actuator. Finally, the output state from the loop interferometer is analysed using pseudo-photon-number resolving techniques via a spatially-multiplexed detector.

% Pulse picking
The pulse-picking system, shown in Fig.~\ref{fig:experiment_setup}(b), sets the clock-time of the experiment and uses a repetition rate of 200 kHz. The pulse configuration of the Pockels cell lets ten pulses of vacuum, ten occupied pulses and then ten more pulses of vacuum. We are interested in those last 20 time-bins, comprising a total time of 250 ns. This means that we are restricting ourselves to an effective duty cycle of $5\%$ of the available pulses. See Appendix~\ref{app:pulse_picker} for further details on this setup.

% losses
The two fields from the SPDC are coupled into single-mode fibres with an efficiency of $\eta_c = 40\%$, without including detection efficiencies. The main limitation here is the mode mismatch between the fibre and the waveguide modes, estimated to be between 60-70\%. We can identify two main sources of loss in the interferometer, the VFBS, with a throughput of $\eta_f = 90\%$ and the loss in the free-space ODL, with an efficiency of $\eta_o = 80\%$. Finally, the detection efficiency of the SNSPDs is maximised by using fibre polarisation controllers to be on average $\eta_d = 80\%$.  

% interference quality 
The purity of the output state, see Appendix \ref{app:source_charac}, is 98\%. Therefore, it can be well approximated by a TMSV state, which can be expressed in the Fock basis as
\begin{equation}
    |\mathrm{TMSV}\rangle = \sqrt{1-\lambda^2}\sum_{n=0}\lambda^n|n, n\rangle,
\end{equation}
where $\lambda = e^{i\theta}\tanh(r)$ and $r$ and $\theta$ are the squeezing and phase parameters, respectively. In the ideal case, interfering the idler and signal modes in a beam splitter gives an SMSV state in each output arm, namely
\begin{equation}
    |\mathrm{SMSV}\rangle = \frac{1}{\sqrt{\cosh(r)}} \sum_{n=0}^{\infty} \frac{\sqrt{(2 n) !}}{2^{n} n !}\left[-e^{i \theta} \tanh (r)\right]^{n}|2 n\rangle
\end{equation}
where the squeezing in each arm is the same as that of the TMSV. The quality of the SMSV will be limited by the interference visibility between the two modes from the TMSV source. To assess this, we perform a Hong-Ou-Mandel measurement between the two fields. From a measurement of the marginal spectrum for each field, see Appendix \ref{app:source_charac}, we observe a slight mismatch in central wavelengths and bandwidth, leading to some distinguishability. The discrepancy in the bandwidth between the two fields originates from the group velocities in ppKTP not being symmetric with respect to that of the pump~\cite{Bell2019}. The Hong-Ou-Mandel measurement reveals a visibility of $95.3(2)\%$, indicating good indistinguishability. We use different squeezing values for the SMSV state, namely $\lambda = \{0.22, 0.31, 0.43\}$.

\begin{figure}[t]
    \centering
    \includegraphics[width=1\columnwidth]{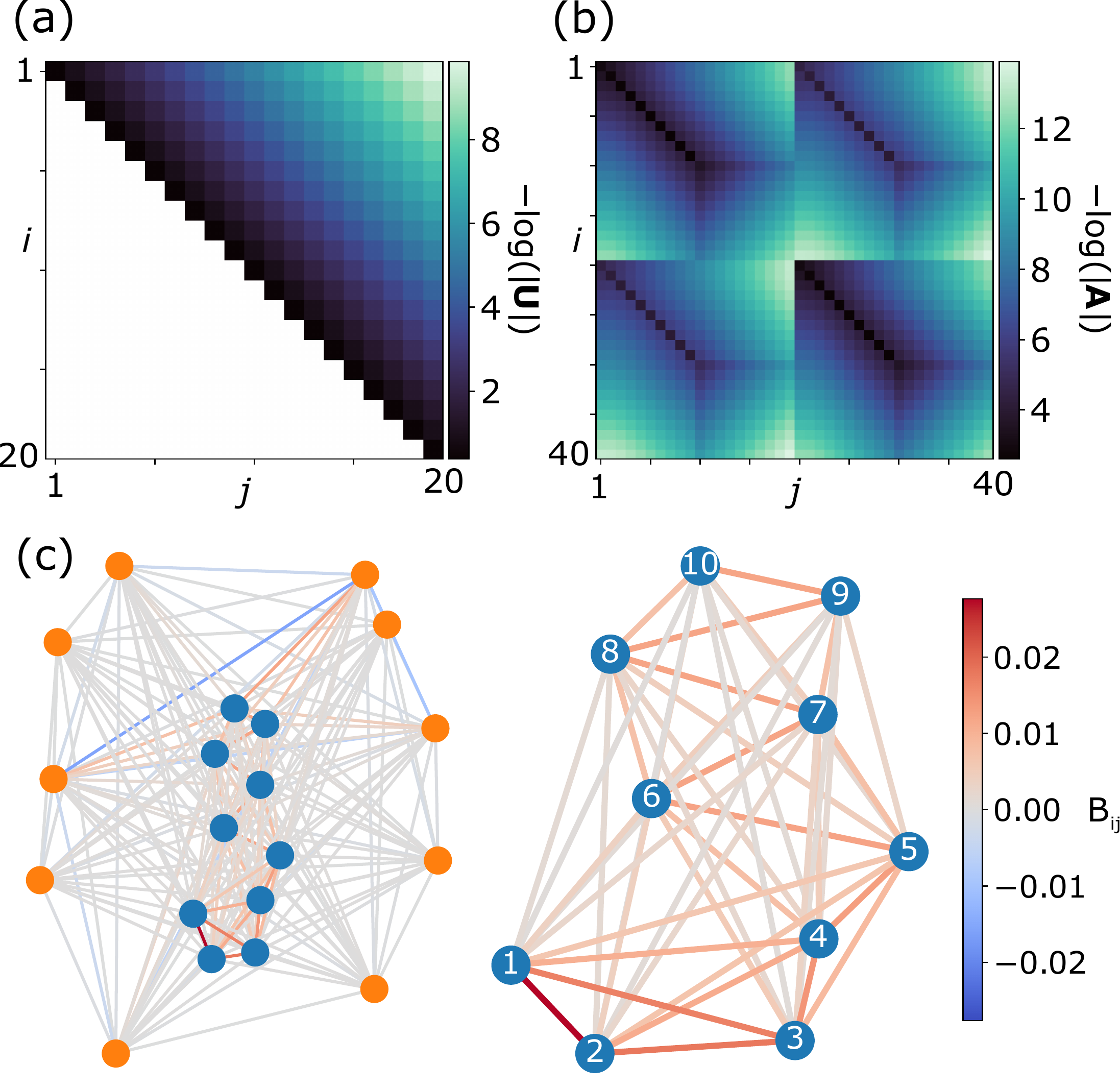}
	\caption{(a) Transformation matrix corresponding to the single loop. (b) $\bm{A}$ of the output state. We can clearly see that the state corresponds to a non-pure Gaussian state, i.e. $\bm{C}\ne0$. (c) (Left) Graph defined by the symmetric part of $\bm{A}$ (top left block matrix) as the adjacency matrix. The blue dots indicate the fully positive subgraph shown on the right. (Right) Fully positive subgraph given by nodes 1-10. The colorbar indicates the weights of the edges.}
	\label{fig:A_mat_phase_locking}
\end{figure}

%interferometer
The single-loop interferometer achieves a sequence of beam-splitting operations between consecutive time-bins, as in Fig.~\ref{fig:circuit}(a) without the SWAP operations. The main motivation behind the use of a single-loop structure, despite its simplicity, is the ability to apply a transformation involving a large number of modes. Fig. \ref{fig:A_mat_phase_locking}(a) shows an estimation of the transfer matrix applied by the single-loop interferometer for $\phi = 0$, where $\phi$ is the phase of the loop. Due to the nature of the single-loop, this is an upper-diagonal matrix since it is not possible to inject a photon at time-bin $i$ and detect it at output $j$ for $i>j$. By varying the reflectivity of the VFBS, we can change the transformation matrix to some extent. In this case, this was arbitrarily set to $T=50\%$. Likewise, Fig.~\ref{fig:A_mat_phase_locking}(b), shows the modelled $\bm{A}$ matrix where, as shown, the $\bm{C}$ matrix is non-zero, denoting the mixedness of the Gaussian state. The piezoelectric actuator allows us to change $\phi$ and perform active phase-locking, see Appendix~\ref{app:phase_locking} for details. Setting $\phi = 0$ makes both the unitary matrix $\bm{U}$ and $\bm{A}$ to be real-valued matrices. %As mentioned above, a particular feature of time-bin interferometers is the non-uniformity of loss across different modes. Usually, in spatial implementations of multiport interferometers loss in different modes can be compensated by adding more loss in the other modes to achieve uniform loss, which can be modelled to happen all at the beginning or at the end of the interferometer. The consequences of not having uniform loss across all the modes are further studied in Sec. \ref{sec:imperfections}.

% PPNR detection scheme
We employ a spatially-multiplexed array of eight superconducting nanowire single-photon detectors to be able to perform pseudo-photon number resolving detection. Due to the reset time of the detectors, $60$ ns, being greater than the spacing between time-bins, 12.5 ns, this type of spatially-multiplexed detection is necessary to be able to resolve detections in consecutive time-bins. This will impact the output detection probability. As an example, take the output state $[1,2]$ (one detection in the first time-bin, and another in the second time-in). This event will be less likely than $[1,10]$ (second detection in the tenth time-bin) since there is a greater chance of the detector being unresponsive due to the detector dead-time. In our case, employing a high number of detectors allows us to neglect this effect.

% figure with U, log(A), phase-locking 

\section{\label{sec:Results}Results}
%We begin by studying the output photon statistics of our device and quantifying the discrepancy between the experimental data and the theoretical model. Then, we use the samples produced by the experimental device to demonstrate an enhancement in the search of dense subgraphs when using these over using uniform random samples.
We begin by analysing the measured photon statistics using a theoretical model of the experiment. Then, we use samples produced by the experimental device to demonstrate an enhancement in the search of dense subgraphs compared to using uniform random samples.
\subsection{\label{sec:gbs_statistics}GBS output statistics}
We model our experiment using the StrawberryFields python package~\cite{KilloranStrawberryComputing}. For this, we use the losses described above, which are measured independently or given by the manufacturer, and assume perfect SMSV input states. The difference between the experiment and theoretical distributions can be computed using the total variation distance (TVD)
\begin{equation}
\label{eq:tvd}
    D(p,q) = \frac{1}{2}\sum_i|p_i-q_i|,
\end{equation}
where $p$ and $q$ are the two probability distributions we are comparing. This distance gives a measure of how close the probability distributions are, ranging from 0 to 1 for identical to completely non-overlapping distributions, respectively. To calculate the probabilities from the model, we use Eq. \eqref{eq:p_n}. 

\begin{figure}[t]
    \centering
    \includegraphics[width=0.9\columnwidth]{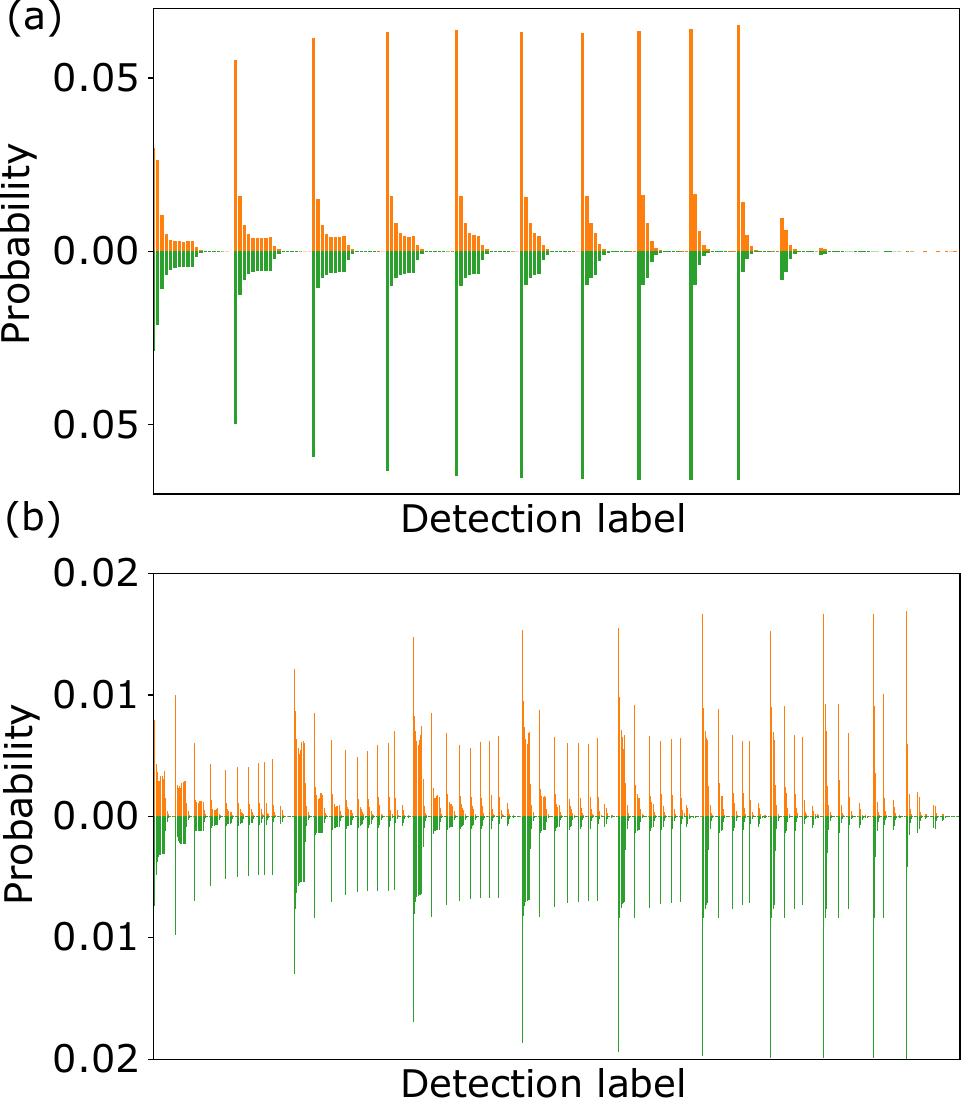}
	\caption{Probability distribution for the twofold (a) and threefold (b) detections with an input squeezing of $\lambda=0.31$. Green bars show the theory values and orange bars the experimental data. The x-axis detection labels are sorted in ascending order from left to right, i.e. $\{[1,1]$, $[1,2]$, \dots, $[19,20]$, $[20,20]\}$ and $\{[1,1, 1]$, $[1,1,2]$, \dots, $[19,20,20]$, $[20,20,20]\}$, respectively.}
	\label{fig:probabiltiy_distribs}
\end{figure}

We minimise the distance for the twofold detections to infer the value of the squeezing at the input. After this, for the $\lambda=0.31$ case, we obtain a distance between the model and the experiment, $D(P_\mathrm{th}, P_\mathrm{exp})$, of 0.0758(3) and 0.081(1), for the twofold and threefold coincidences, respectively, indicating the model does a good job in describing the experiment.  Fig.~\ref{fig:probabiltiy_distribs} shows a plot with the (a) twofold and (b) threefold probabilities obtained from the experiment and given by the model. We observe a periodic pattern depicting several exponential decays, which is characteristic of the single loop transformation. In both cases, we see good agreement between the theory and the experiment. The detection rates of threefold and fourfold detections are 300 Hz and 10 Hz, respectively. Note that the experiment rate is 200 kHz with a duty cycle of 5\%, limited by the Pockels cell. By overcoming these limitations, these rates could be increased by a factor of 20 for the same level of squeezing.
\subsection{\label{sec:finding_subgraphs_results}Finding dense subgraphs}
%Here we experimentally demonstrate that a GBS device can be used to enhance the search for dense subgraphs in a graph.
As mentioned earlier, in a GBS experiment the graph we want to investigate is encoded in the symmetric part of $\bm{A}$, i.e. the $\bm{B}$ matrix. For the purpose of identifying dense subgraphs, we require the adjacency matrix of the graph to be non-negative or non-positive, such that the hafnian is correlated with the density. For this, we focus on a subgraph of the graph defined by $\bm{B}$ for the $\lambda = 0.31$ case that solely contains positive weights. We choose the subgraph containing nodes 1-10, which corresponds to time-bins 1-10. Fig. \ref{fig:A_mat_phase_locking}(c) shows both the complete graph, given by $\bm{B}$, (left) and the fully positive subgraph defined by nodes 1-10 (right). Since we are interested in $k$-node subgraphs, we filter the obtained samples to only select nondegenerate $k$-fold detections. In this particular case, due to the nature of the interferometer, this has a substantial impact on the detection rates. After this postselection, we are left with 107, 312 and 1802 nondegenerate fourfold samples for $\lambda = \{0.22, 0.31, 0.43\}$ with integration times of 8h, 3h and 3h, respectively. 

%Similar to~\cite{arrazola2018dense}, we take the postselected GBS samples and use them as seeds in a classical algorithm to search for dense subgraphs, in this case, a random search. In the random search algorithm, we draw $n$ samples, each containing $k$ nodes, from the appropriate distribution, calculate the density corresponding to each sample, and select the one with the maximum density. We vary the number of samples $n$ drawn and, for each of these values, repeat the procedure 400 times to remove statistical fluctuations.
The use of GBS samples to identify dense subgraphs is demonstrated with a classical random search algorithm, as in~\cite{arrazola2018dense}. In the random search algorithm, we draw $n$ samples, each containing $k$ nodes, calculate the density corresponding to each sample, and select the one with the maximum density. We vary the number of samples $n$ drawn and, for each of these values, repeat the procedure 400 times and calculate the mean to remove statistical fluctuations.

\begin{figure}[ht]
    \centering
    \includegraphics[width=1\columnwidth]{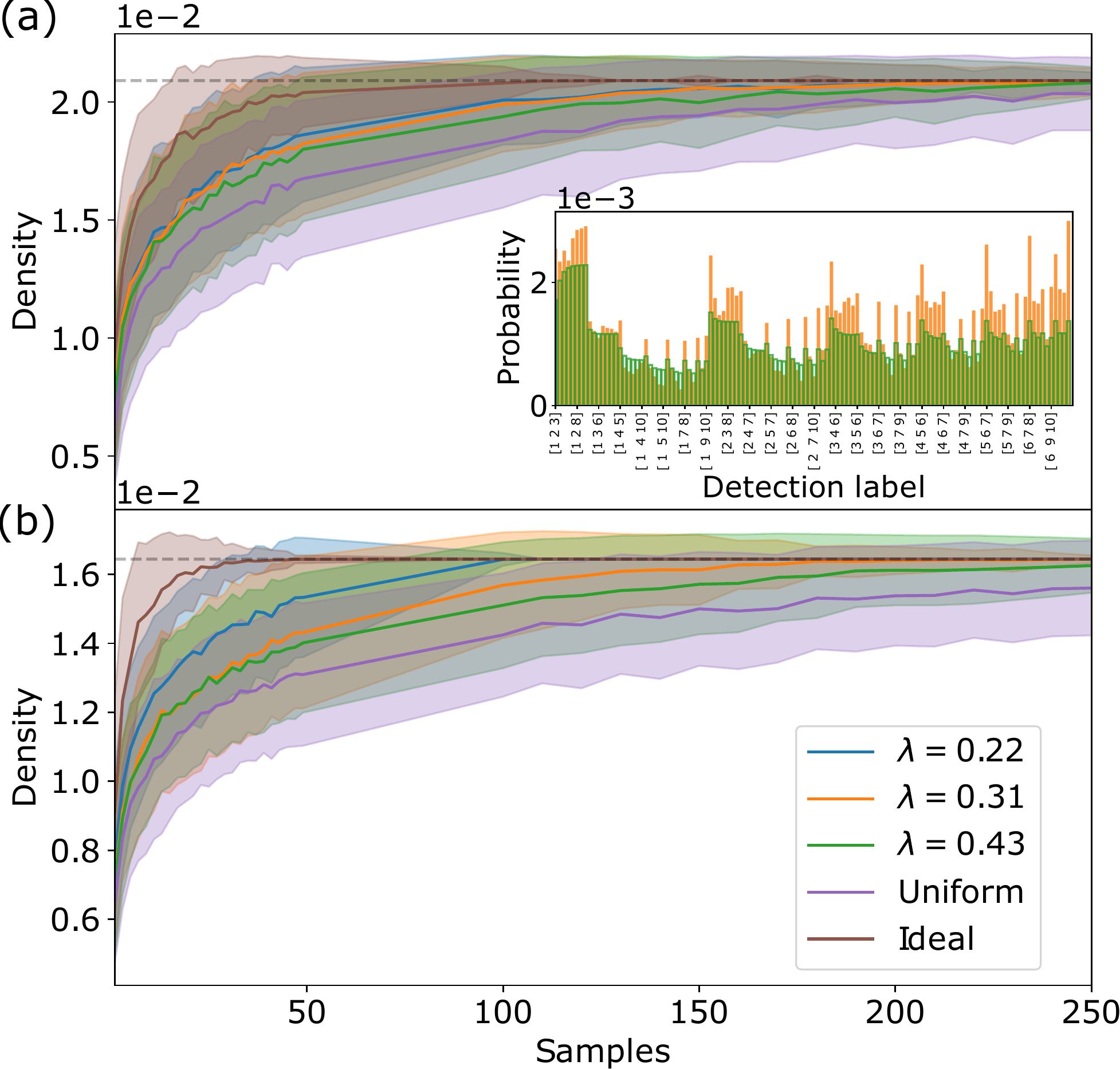}
	\caption{(a) Mean density of the 3-node subgraph as a function of the number of samples drawn. The inset shows the probability distribution for the nondegenerate output states of the experimental samples (orange filling) and the theoretical model (green edge). (b) Mean density of the 4-node subgraph as a function of the number of samples drawn. The purple line indicates a uniform random search, the blue, orange and green lines use samples obtained from the GBS experiment for different values of the squeezing as indicated in the legend, and the brown line indicates samples obtained when sampling from the distribution obtained from $|\mathrm{Haf}(\bm{B_n})|^2$. The dashed line indicates the density of the densest subgraph. The search speed decreases when increasing the squeezing. This effect is further studied in Sec.~\ref{sec:imperfections}.}
	\label{fig:density_vs_samples}
\end{figure}

Fig.~\ref{fig:density_vs_samples} shows the mean density obtained as a function of the number of samples drawn for subgraphs of size three (a) and size four (b). We see that in both cases the GBS-enhanced protocols perform better than the uniform sampling case, this improvement being more prominent in the 4-node subgraph case. To compare the performance of the algorithm for the different input seeds, we consider the number of samples needed, on average, to find a graph whose density is 95\% of the maximum and the mean density achieved for a given number of samples. The results are summarised in Table~\ref{tab:samples_at_90} and give a quantitative value of the speedup provided by the GBS device.

The ideal curve corresponds to the loss-free system. In this case, the samples are drawn directly from the distribution given by $|\mathrm{Haf}(\bm{B_n})|^2$. For the threefold case, where the hafnian would be null due to the number of detections being odd, we use the fourfold distribution and remove one detection at random. As observed, the performance of the algorithm using these samples is the best and considerably better than the uniform sampling case. Then, imperfections in the experimental setup mean that we do not sample anymore from $\bm{B}$ [Eq.~\eqref{eq:haf_B}] but need to use $\bm{A}$ [Eq.~\eqref{eq:haf}], as $\bm{C}\ne0$. Crucially, we find that despite significant optical loss the performance of the algorithm when using experimental samples still notably surpasses that of the uniform sampling case. Interestingly, we also notice that when increasing the squeezing the speed at which the curve approaches the maximum mean density decreases. Table~\ref{tab:samples_at_90} quantitatively illustrates this observation. This effect is studied in more detail in Section~\ref{sec:imperfections}. 
%Moreover, as observed, imperfections in the experiment, mostly due to optical loss, degrade the performance of the algorithm, as indicated by the ideal curve in (b), which would correspond to the lossless scenario. In this case, the samples are drawn directly from the distribution given by $|\mathrm{Haf}(\bm{B_n})|^2$. For the threefold case, where the hafnian would be null due to the number of detections being odd, we use the fourfold distribution and remove one detection at random. Interestingly, we also notice that when increasing the squeezing the speed at which the curve approaches the maximum mean density decreases. Table~\ref{tab:samples_at_90} quantitatively illustrates this observation. This effect is studied in more detail in Section~\ref{sec:imperfections}. 
%lossy case
\begin{table}[h]
\caption{\label{tab:samples_at_90} Subgraph search performance for different sources of random search seeds. The metrics shown are: the number of samples needed to obtain a mean density of 95\% and the density achieved when using 50 samples.
}
\begin{ruledtabular}
\begin{tabular}{cccc}
$n$ & Seed & Samples at 95\% den. & Density for 50 samp.\\
\colrule
3 & Ideal & 34(1) & 0.020(0)\\
3 & $\lambda = 0.22$ & 92(3) & 0.0186(1)\\
3 & $\lambda = 0.31$ & 99(3) & 0.0182(1)\\
3 & $\lambda = 0.43$ & 117(4) & 0.0180(2)\\
3 & Uniform & 178(5) & 0.0167(2)\\[0.1cm]
4 & Ideal & 14(1) & 0.0164(0)\\
4 & $\lambda = 0.22$ & 62(3) & 0.0153(0)\\
4 & $\lambda = 0.31$ & 98(3) & 0.0143(1)\\
4 & $\lambda = 0.43$ & 143(5) & 0.0140(1)\\
4 & Uniform & $>260$ & 0.0130(1)\\
\end{tabular}
\end{ruledtabular}
\end{table}

\section{\label{sec:imperfections}Loss and squeezing trade-off}
\begin{figure*}[t]
    \centering
    \includegraphics[width=1\textwidth]{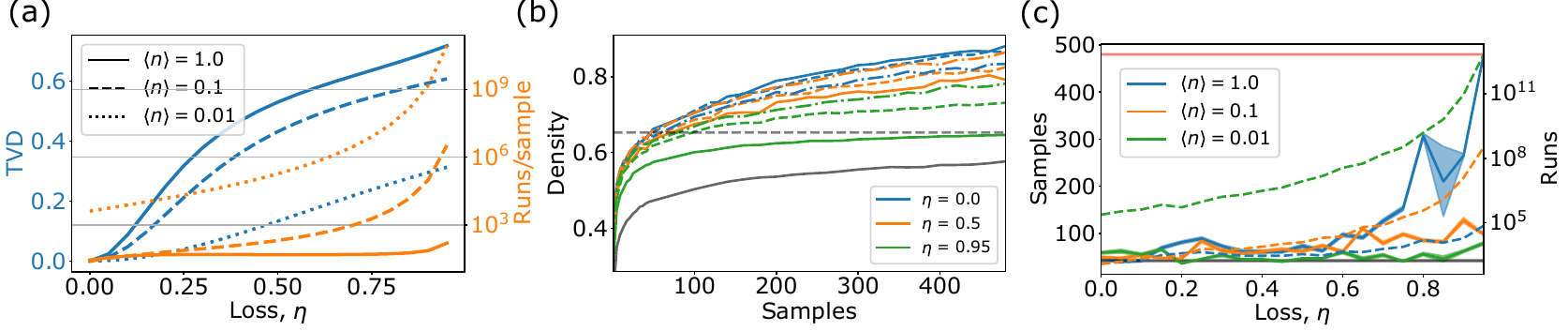}
	\caption{(a) TVD and runs/samples as a function of the input loss for several mean photon numbers per mode at the output. (b) Density as a function of the number of samples used in the random search algorithm for several mean photon numbers indicated in the legend of (a). The different colours correspond to different input losses in the system while the black curve shows the performance of the algorithm when using uniform random samples. (c) Number of samples (solid) and runs (dashed) needed to achieve a certain mean density, defined by the black dashed line in (b), as a function of the input loss for several mean photon numbers. The shaded areas correspond to one standard deviation of the mean. The black line indicates the density achieved by the algorithm when using samples drawn directly from the adjacency matrix of the graph and the red line corresponds to uniform drawn samples. The shaded areas correspond to one standard deviation of the mean.  %(d) Mean density (solid) and runs (dashed) achieved when using 300 samples, indicated by the dotted line in (b), as a function of the input loss for several mean photon numbers. 
	}
	\label{fig:imperfections_plot}
\end{figure*}

In this section, we study how imperfections affect GBS outputs and, consequently, impact the performance of the classical algorithm to find dense subgraphs. In essence, these imperfections will change the probability distribution we are drawing our samples from, but how this modifies the performance of the algorithm for this task is not clear. %Previous work~\cite{Su2021} has focused on methods to mitigate the effect of loss in the probability distribution, although for large systems performing the additional measurements required for this strategy may prove challenging. 
Here, we focus on the most prominent imperfection in typical GBS experiments: uniform channel loss, for example due to source-interferometer coupling or detector inefficiencies. We acknowledge that other errors such as deviations from the expected unitary or mode-dependent losses are also possible, but we leave these for future work.

To study the role of imperfections we use the following procedure
\begin{enumerate}
    \item Construct the problem graph by generating two unweighted Erdos-Renyi graphs, a small one with high edge probability and a larger one with low edge probability. Connect the nodes of the smaller graph to the nodes of the larger one at random~\cite{arrazola2018dense}.
    \item  Perform a Takagi-Autonne decomposition to obtain the squeezing and interferometer parameters that encode this graph in a GBS device. Rescale the squeezing values to change the mean photon number per mode at the output.
    %\item Modify the circuit appropriately, e.g. adding optical loss at the modes, to effectively study how the imperfections affect the output probabilities. We investigate how this probability distribution deviates from the lossless case.
    \item Add identical optical loss to each mode of GBS device model.
    \item Using the graph and samples from the GBS model, a subgraph search is conducted as in the experiment sections above.
    % \item Generate samples according to those probabilities and feed them into the random search algorithm previously described to find dense subgraphs. Repeat this for a different number of input samples. 
    % \item Find the number of samples needed to achieve a mean density over the $n$ repetitions to be above a certain threshold and the density achieved with some number of samples.
\end{enumerate}
Here, we present results for a graph consisting of 26 nodes, where a 6-node graph with edge probability 0.875 is joined to a 20-node graph with edge probability 0.3. For the random search algorithm, we repeat the search for $1000$ iterations to precisely determine the mean performance.

First, we investigate how loss modifies the normalised probability distribution for a given $n$-fold detection. To evaluate this, we use the TVD, introduced in Eq. \eqref{eq:tvd}. In other types of experiments, where we have a well-defined number of input photons, e.g. boson sampling, these probability distributions remain independent of loss. Fig.~\ref{fig:imperfections_plot}(a) shows the TVD with respect to the lossless case and the average number of runs needed to obtain a $6$-fold sample, i.e. $1/p(6)$ where $p(6)$ is the probability to obtain a collision-free 6-fold detection at the output, as a function of the loss for several mean photon numbers per mode, i.e. $\langle n \rangle = 1/m \sum_i \langle n_i \rangle$ where $m$ is the number of modes. We observe that as the loss increases, the deviation from the lossless case becomes more pronounced, particularly for those cases in which the squeezing was high. In the low squeezing limit, the probability of generating instances where more photons are generated than detected, i.e. losing photons, is minimal, leading to a vanishing TVD.
%When the squeezing is low, we can truncate the superposition of even number Fock states at $|n\rangle \rightarrow 2$, in the low squeezing limit, and, hence, the impact of higher-order photon numbers being affected by loss is negligible.

Considering the number of GBS runs required for a $6$-fold detection event, runs/sample from hereon, we see that the choice of squeezing presents a tradeoff between TVD and detection rates. We proceed to study how this compromise manifests in the search for dense subgraphs and if, at some point, allowing more runs/samples can lead to a substantial decrease in the TVD that is beneficial to this problem. An alternative to reducing the squeezing may be error mitigation~\cite{Su2021}. This procedure involves either changing the experimental loss parameter to interpolate the measurements or performing classical postprocessing of the data. Practically, these techniques may prove challenging when studying larger systems.

%Similar to Fig.~\ref{fig:density_vs_samples}, Fig.~\ref{fig:imperfections_plot}(b) shows the density obtained with $n$ samples using a random search algorithm for several values of input loss (different line colours) and different mean photon numbers (different line styles). 
The mean density obtained with $n$ samples using a random search algorithm for several values of input loss and squeezing parameters, indicated by the corresponding line colours and styles, respectively, is shown in Fig.~\ref{fig:density_vs_samples}(b). %The grey dashed line indicates the value for the density we use to calculate the data shown in Fig.~\ref{fig:imperfections_plot}(c). 
As observed, for a constant value of the loss, the performance of the algorithm diminishes with increasing values of squeezing, in good agreement with the experimental observations. Fig.~\ref{fig:imperfections_plot}(c) shows the point where the mean density achieved by each of the curves in (b) crosses the dashed line, i.e. achieves a mean value of 75 \%, as a function of loss for different mean photon numbers. The speed at which the algorithm reaches a certain value for the mean density degrades as a function of the input loss for high squeezing values, but it remains unaffected when the squeezing is small. %Likewise, Fig.~\ref{fig:imperfections_plot}(d) shows the crossing points with the dotted line in (b) as a function of loss for several mean photon numbers. Similarly, the value achieved after the rapid growth in density at the beginning also changes depending on the squeezing applied, as shown in Fig.~\ref{fig:imperfections_plot}(d).
Note that in all cases, the GBS device still outperforms the uniform sampling approach indicated by the solid grey line. The dashed lines indicate the number of experimental runs needed to achieve the required number of samples needed to obtain a mean density. As shown, because the number of runs/sample increases exponentially with the mean photon number $\langle n \rangle$, in the regime we have studied, increasing the squeezing reduces the number of runs needed, in general. Due to the statistical noise of the algorithm, it is hard to extract conclusions for regimes where the variations between average photon numbers are small. Despite the success of our lossy experiment, these results indicate that for large-scale dense subgraph searches, where high squeezing values are required, further consideration of the impact of loss is likely needed. In some instances, it might even be advantageous to decrease the squeezing to reduce the TVD, despite the corresponding sacrifice in detection rates.

\section{\label{sec:conclusions}Conclusions}
%summary of the work

We have demonstrated that measurements from a realistic GBS device can enhance the search for dense subgraphs over uniform random sampling techniques. To do so, we demonstrate GBS in a time-bin implementation with a single source of SMSV. We study up to four-photon detection events over twenty modes, showing that a single-loop interferometer can readily scale to tens of modes for this application. Then, we map the output Gaussian state to a graph and choose a subset of detection outcomes corresponding to a positive subgraph. We show that any specific outcome occurs with a probability dependent on the density of its subgraph and, therefore, demonstrate that GBS can enhance a search algorithm, e.g. random search, for the identification of dense subgraphs of a graph. %(Add comment about subset of detection outcomes corresponding to a positive graph, with any specific outcome occuring with a probability dependent on the density of a subgraph.)

% To summarise, we used a time-bin architecture to experimentally demonstrate that samples from a GBS device can be used to enhance the search for dense subgraphs over uniform random sampling techniques. We employed a time-bin interferometer and a source of SMSV to detect samples with up to four photons. The use of a time-bin configuration as opposed to a spatial mode interferometer allowed us to scale up the number of modes in a straightforward manner. From the generated symmetric $\bm{B}$ matrix, we identified a fully positive submatrix and mapped it into a graph through its adjacency matrix. By post selecting samples contained in the rows and columns of such submatrix, we showed that when feeding the random search algorithm with those samples, the speed at which we can find dense subgraphs is enhanced over uniform random samples. 

We repeat this procedure for three different squeezing parameters, observing a degradation of the algorithm's performance with increasing squeezing. To understand this, we numerically studied the role of input loss in a GBS experiment and found that the TVD with respect to the lossless case increases when using a higher squeezing. We then investigated the impact of this in the search for dense subgraphs problem and showed that there exists a trade-off between squeezing and algorithm speed for high values of input loss. This is in good agreement with the experimental observations. %Due to the different scalings between the degradation of the speed of the algorithm and the number of samples obtained per run as a function of the squeezing, one being polynomial while the other showing an exponential dependence, we do not expect this to play a relevant role for the time being.

% relevance of this work
This work constitutes the first experimental demonstration of the implementation of GBS in a time-bin encoded architecture and of how a GBS device can be employed to speed up the search of dense subgraphs. We hope this work can motivate other research groups to scale up this application to the regime where using a GBS device gives a quantum advantage. In this sense,  in-depth consideration of what the consequences of imperfections are in near-term applications using GBS devices is needed.

Time-bin encoded GBS, in a fibre-based interferometer, offers many practical advantages for achieving quantum speedups in certain application areas. Our work should encourage further theoretical studies of computational problems which map onto this architecture.  It will stimulate the engineering of integrated fibre-based squeezed light sources~\cite{Lugani2020a} as well as low-loss switching, which is essential for performing arbitrary operations on time-bin encoded states.  

Lastly, we note that as we were making final edits to our manuscript, a closely related paper was published~\cite{Madsen2022}. This work exemplifies the great potential of time-bin encoded architectures to scale up GBS experiments to a regime where they are no longer classically simulable. Our work provides a route to extend this approach to obtain a quantum advantage in a relevant application, such as the search for dense subgraphs, under realistic experimental conditions.

\begin{acknowledgements}
We thank Jamie Francis-Jones for valuable input at the initial stage of this project and Guillaume Thekkadath for insightful discussions. This work was supported by: Engineering and Physical Sciences Research Council
via the Quantum Systems Engineering Skills Hub and the Quantum Computing and Simulation Hub (P510257, T001062) 
\end{acknowledgements}

\appendix

\section{Pulse picker}
\label{app:pulse_picker}
As shown in Fig.~\ref{fig:experiment_setup}(b), as the beam enters the pulse-picking sytem, it first passes through a Glan-Taylor (GT) polariser with its polarization axis aligned to the input light. A half-wave plate (HWP) is then used to rotate the polarization of the beam to 45$^{\circ}$ with respect to the optic-axis of the Pockels cell. The Pockels cell consists of an X-cut 20 mm rubydium-tantalate phospate (RTP) crystal with an aperture of 3~mm and $V_\pi = 1$~kV. Its high-voltage driver is triggered by a digital delay generator at a repetition rate of 200~kHz. A quarter-wave plate (QWP) compensates for the natural birefrigence of the RTP crystal followed by a second GT polariser. The system is arranged in a double-pass configuration to further increase the extinction ratio, which we observe to be $1:10^5$. The overall transmission through the system is 70\%. Fig.~\ref{fig:spectrum_HOM}(c) shows a histogram of the counts after the pulse picker. For the data acquisition, we gate the timetagger system to only record time tags in the [0, 250] ns region, reducing the required analysis time substantially.

\section{Source characterization}
\label{app:source_charac}
\begin{figure}[t]
    \centering
    \includegraphics[width = \columnwidth]{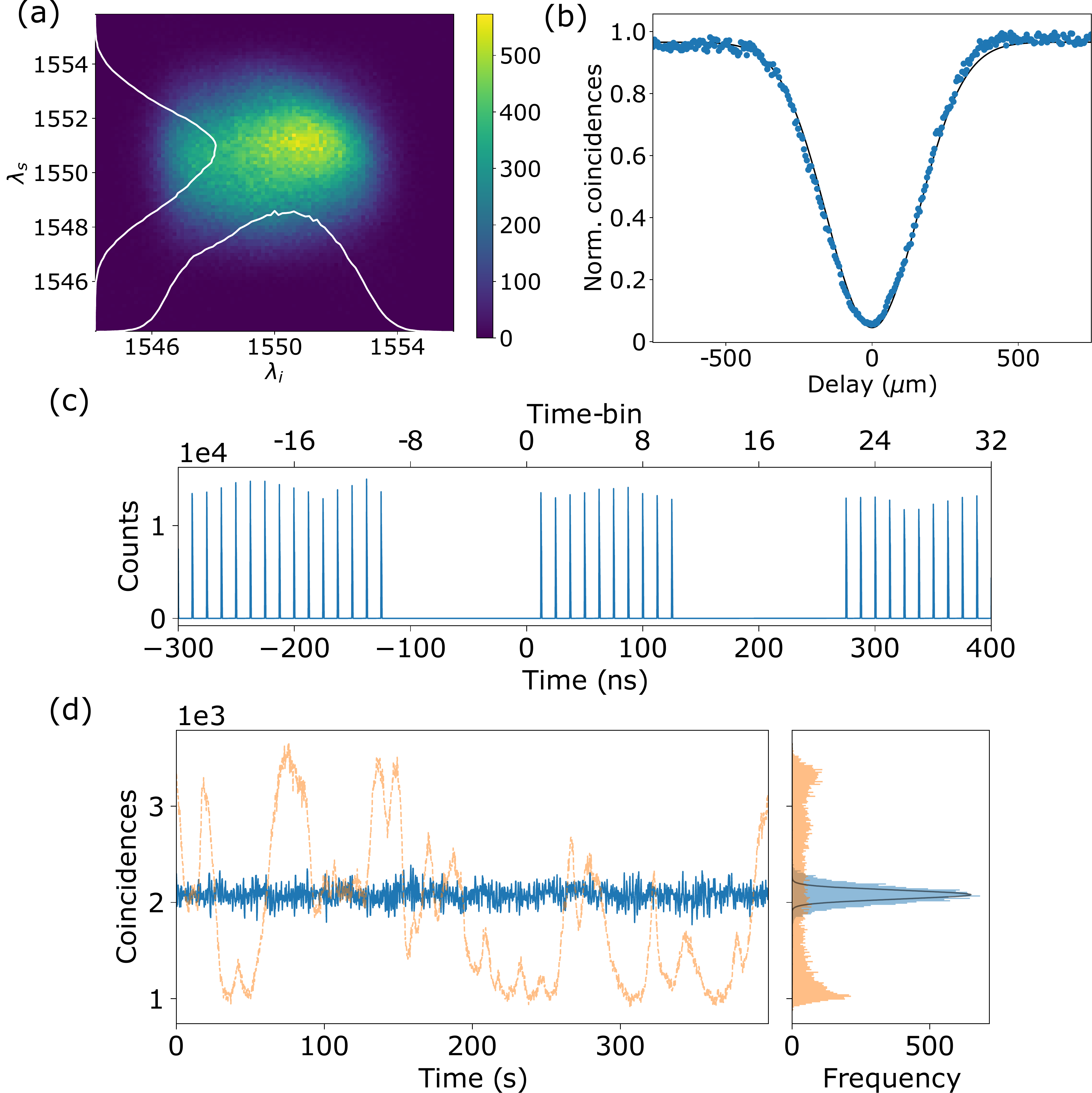}
    \caption{(a) Joint spectral intensity of the source, with the number of counts indicated by color. (b) Hong-Ou-Mandel interference with a visibility of $95.3(2)\%$. The error bars, due to Poissonian counting statistics, are barely visible. (c) Timing histogram of counts at the unused output port of the FBS, demonstrating pulse picking. (d) Twofold coincidences after the interferometer for a freely running (orange) and locked (blue) loop phase. The black line in the histogram indicates fluctuations due to Poissonian statistics.}
    \label{fig:spectrum_HOM}
\end{figure}

The joint spectral intensity of the signal and idler fields generated by SPDC, shown in Fig.~\ref{fig:spectrum_HOM}(a), is measured with time-of-flight spectrometers that employ dispersion-compensating fibers followed by single-photon detectors. Assuming a uniform joint spectral phase, this measurement indicates an effective mode number of $1.02$. To assess the degree to which the marginal signal and idler spectra are identical, we study Hong-Ou-Mandel interference by measuring coincident photons at the output of the beam splitter (FBS) in Fig.~\ref{fig:experiment_setup} when working in the low-squeezing regime. We obtain a visibility of $95.3(2) \%$, defined as $(C_{\max}-C_{\min})/C_{\max}$. 

\section{Phase locking}
\label{app:phase_locking}
The round-trip phase of the interferometer loop is adjusted using piezoelectric control of a mirror position. We employ a piezoeletric stack with a no-load maximum displacement of 12 $\mu$m glued to a 1/2-inch mirror. A control signal is derived from the rate at which two coincident photons are observed at the output of the interferometer. The corresponding detection events are integrated for 150~ms and used for PID control of a piezo signal in the range of 0-7~V. Fig.~\ref{fig:A_mat_phase_locking}(d) compares the coincidence detection events when the loop runs freely and when it is locked.

%\section{Imperfections}
%\label{app:imperfections}
%Here we extend the exploration of imperfections in graphs to 
%multiple sizes. Table~\ref{tab:graph_params} shows the parameters used to construct each of these graphs. 
%\begin{table}[b]%The best place to locate the table environment is directly after its first reference in text
%\caption{\label{tab:graph_params}%
%}
%\begin{ruledtabular}
%\begin{tabular}{cccc}
%$N_A$ & $N_B$ & $p_A$ & $p_B$\\
%\colrule
%4 & 12 & 0.3 & 0.875\\
%6 & 12 & 0.3 & 0.875\\
%6 & 26 & 0.3 & 0.875\\
%\end{tabular}
%\end{ruledtabular}
%\end{table}
\bibliographystyle{apsrev4-2}
\bibliography{GBS_dense_subgraphs}

\end{document}